\documentclass[12pt]{article}
\usepackage[dvips]{graphicx}
\newcommand{\be}{\begin{equation}}
\newcommand{\ba}{\begin{array}}
\newcommand{\bd}{\begin{displaymath}}
\newcommand{\ee}{\end{equation}}
\newcommand{\ea}{\end{array}}
\newcommand{\ed}{\end{displaymath}}

\def\lsim{\mathrel{\rlap{\lower4pt\hbox{\hskip1pt$\sim$}}
    \raise1pt\hbox{$<$}}}      
\def\gsim{\mathrel{\rlap{\lower4pt\hbox{\hskip1pt$\sim$}}
    \raise1pt\hbox{$>$}}}      

\def\frac#1#2{{#1\over #2}}

\def\xi2{$\chi^2_{d.o.f}$}
\def\x2{$\chi^2$}

\def\g2{ GeV$^2$}
\def\G2{{\rm (GeV}^2)}

\def\ie{\hbox{\it i.e. }}
\def\etc{\hbox{\it etc... }}
\def\eg{\hbox{\it e.g. }}

\def\etal{\hbox{\it et al. }}

\begin{document}

\bigskip
\rightline{LYCEN 2000-93}
\rightline{September 2000}

\bigskip
\bigskip

\begin{center}
{\Large \bf Damping of the HERA effect in DIS~?}
\bigskip
\bigskip
\end{center}

\begin{center}

{\large {
P. Desgrolard $^{a,}$\footnote{E-mail:
desgrolard@ipnl.in2p3.fr},
A. Lengyel $^{b,}$\footnote{E-mail:
sasha@len.uzhgorod.ua},
E. Martynov $^{c,}$\footnote{E-mail:
martynov@bitp.kiev.ua} }}

\end{center}

\bigskip
\bigskip

\noindent
$^a$ Institut de Physique Nucl\'eaire de Lyon, IN2P3-CNRS et
Universit\'e Claude Bernard, 43 boulevard du 11 novembre 1918, F-69622
Villeurbanne Cedex, France

\noindent $^b$ Institute of Electron Physics, National Academy of Sciences
of Ukraine, 08015 Uzhgorod-015, Universitetska 21, Ukraine

\noindent $^c$ Bogoliubov Institute for Theoretical Physics, National
Academy of Sciences of Ukraine, 03143 Kiev-143, Metrologicheskaja 14b,
Ukraine

\bigskip
\bigskip

\begin{center}
\begin{minipage}[t]{13.0cm}
\noindent
{\bf Abstract}
The drastic rise of the proton structure function $ F_2(x,Q^2)$
 when the Bj\"orken variable
$x$ decreases, seen at HERA for a large span of $Q^2$, negative
values for the 4-momentum transfer, may be damped when $Q^2$
increases beyond $\sim $ several hundreds \g2. A new data
analysis and a comparison with recent models for the proton
structure function is proposed to discuss this phenomenon in terms
of the derivative $\partial{\ell n F_2(x,Q^2)}/\partial{\ell
n(1/x)}$.
\end{minipage}
\end{center}


\section{\Large{Introduction}}
The so-called "HERA effect" discovered some years ago by the
experimentalists in deep inelastic scattering (DIS) and
anticipated by a number of theoreticians has arisen a great
interest in the community (see {\it e.g.}\cite{Lev} and references
therein). It concerns the strong rise of the proton structure
functions (SF) $F_2(x,Q^2)$ when the Bj\"orken variable $x$
decreases, for the experimentally investigated $Q^2$, negative
values for the 4-momentum transfer.

A slow-down of a further rise is inevitable, as a consequence of
unitarity, if however the Froissart-Martin \cite{F-M} bound is
valid for $\gamma^{*} p$ interaction at least at $x\to 0$. In this
case a power-like behavior of SF at small-$x$ must be transformed
into a logarithmic-like one (like in hadron amplitudes, when a
procedure of unitarization or eikonalization is applied).
Evidently, such an evolution leads to a damping of the fast growth
of SF at $x\to 0$.

A natural question arises~: will this HERA effect still subsist in
any $x-Q^2$ region to be investigated, or more pragmatically in
which kinematical region a damping is to be expected~?

It is not easy to exhibit this interesting tendency in the
behavior of $F_2(x,Q^2)$ at small $x$ and rising $Q^2$. This
effect is very weak because a sufficient amount of data at high
$Q^2$ is still lacking. Furthermore, it may stand near the
kinematical limit at the available energies (it is the case at
HERA~\footnote{the kinematical limit $y ={Q^2\over x(s-m_p^2)}\le
1$ for HERA measurements, with $s-m_p^2\approx 4E_e E_p$, in terms
of the positron $E_e$ and proton $E_p$ beam energies, writes
$Q^2$(\g2 ) $\lsim 10^5 x$ since August 1998.}), thus preventing a
clear evidence on the basis of existing data. Nevertheless, one
can guess it by simply inspecting by eye the evolution of the
icons showing the experimental SF versus $x$ at given $Q^2$
increasing up to the highest values (see in Fig.1 such a
representation using a double logarithmic scale); if this
inspection seems inconclusive, one can at least insure that no
contradiction appears.
\begin{figure}[ht]\label{f.1}
\begin{center}
\includegraphics*[scale=0.7]{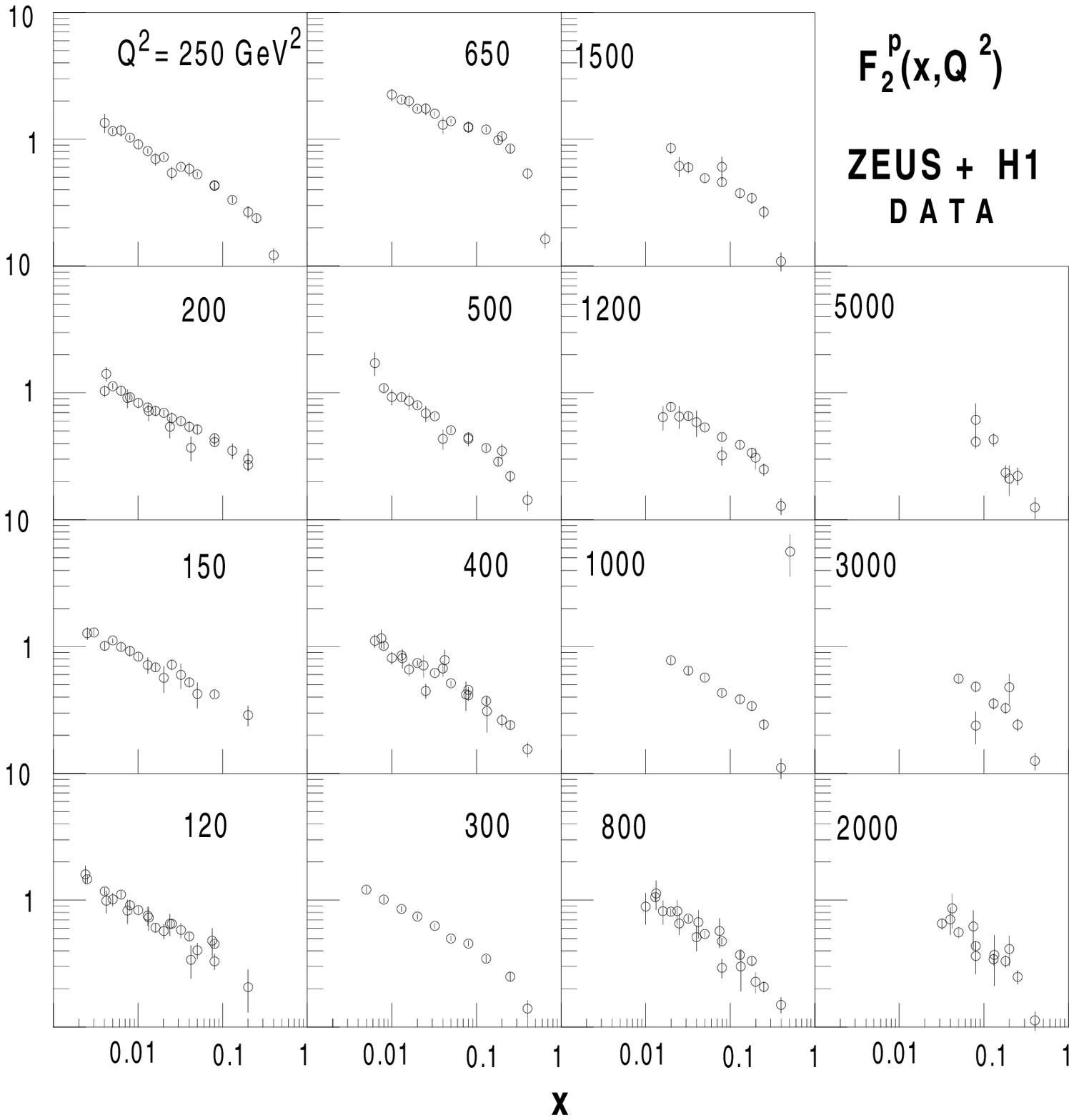}
\caption{Experimental data of ZEUS and H1 collaborations at high
$Q^{2}$}
\end{center}
\end{figure}

Aside from this empirical mean, a way to quantify the preceding
effect is to consider the derivative of the logarithm of the
proton SF with respect to $\ell n(1/x)$ ($x-$slope for brevity)
defined as
\be B_x(x, Q^2) = {\partial\ell n F_2(x,Q^2)\over
\partial{\ell n (1/x)}}\ ;
\ee
the damping of the HERA effect would correspond to the presence of
a maximum when plotting $B_x$ versus $Q^2$ for a given low $x$.

It is worth noting that this derivative is
related to an "effective power" (exponent of $1/x$) $\Delta$,
sometimes labelled $\lambda$, introduced when redefining the
proton SF as \be
 F_2 (x,Q^2)= G(Q^2)\ \left({1\over x}\right)^{\Delta (x,Q^2)}\ ,
\ee or in other terms to a Pomeron "effective intercept"
$\alpha_{\cal P}^{eff}$ (depending on $x$ and $Q^2$) \be
\alpha_{\cal P}^{eff}(x,Q^2) = 1 + \Delta(x, Q^2) \ . \ee We
emphasize that we can identify $ B_x =\Delta$, only in the case of
a model where $\Delta$ is $x$-independent; in general $B_x$ and
$\Delta$ do not coincide.
A careful analysis of experimental data, based on a comparison of
various parame\-trizations of the SF, has suggested~\cite{djlp} a
possible slowing down of the SF rise revealed as a growth of $F_2$
with decreasing $x$ becoming steeper and steeper (when $Q^2$
rises) which is changed at $Q^2\sim 200-500$ GeV$^2$ into a growth
saturating and even becoming less and less steep at further
increasing $Q^2$.

\bigskip

From a phenomenological point of view, many parametrizations of
the proton structure functions successfully accommodated for the
whole - or a part of - available data set and in particular for the
steep rise of the SF when $x$ decreases for a large span of $Q^2$
values (see \eg ~[4-17]). Several of them have given explicitly
some hints (see \eg~\cite{Be-95,djp,dlm}) on the existence of a
slowing down of the HERA effect.

We believe that the region of the contour $x-Q^2$ plane with
$x\lsim 10^{-1}$ and $Q^2\gsim 100$ \g2 is of particular interest
for the above question. Happily, recent measurements of SF have
become available, from H1 and ZEUS collaborations which fulfil
partly this kinematical criterion. They complete or correct the
previous data near the HERA collider~\cite{ah95,ai96,ad97,ad99}
and~\cite{de96,br97,br99,br00} and from fixed target
experiments~\cite{ar97,ad96,slac,be89,ca75}.

Here, we actualize the conclusions of previous paths driving towards the
existence of a damping effect. We perform with new data a relevant
analysis similar to~\cite{djlp}, where, however, only the averaged slopes
$B_{x}(x,Q^{2})$ were determined in almost the whole available interval of
$x$ at given fixed $Q^{2}$.

It is necessary to remark here that the data on $\lambda_{eff}$ \cite{br99}
cannot be interpreted as showing a dependence of the effective
intercept ($\Delta_{eff}(Q^{2})$) on $Q^{2}$. Each point $B_{x}$ has not been
extracted at the same $<x>$, There is an evident correlation between the
values of $Q^{2}$ and of the corresponding $<x>$, namely a larger $<x>$
corresponds to a larger $Q^{2}$ for almost all points \footnote{See also a
discussion about $\Delta_{eff}$ in the second paper of \cite{CKMT}.}.
To exhibit a true dependence of the effective intercept on $Q^{2}$, one need
to have the local $B_{x}$ extracted at the same $<x>$ but for different
$Q^{2}$.
It is one of the reasons motivating the present work.
We attempt a new, more detailed, systematic (on all available data) and
model independent analysis of the local slope $B_{x}(x,Q^{2})$ having in
mind its possible variation with $x$. It allows not only to extract the
averaged slopes at fixed $Q^{2}$ but also to determine the dependence of
$B_{x}(x,Q^{2})$ on $Q^{2}$ at fixed $x$.

We also turn towards three basically different phenomenological
models for the proton SF, namely: the "ALLM" model~\cite{AL-97},
the "interpolating Lyon-Kiev-Padova" (LKP) model~\cite{djp},
the "soft dipole Pomeron" (DP) model~\cite{dlm},
and  have been refitted (or eventually improved as concerns LKP) with a
common set of data including the most recent ones, their predictions
concerning the $x-$slope are compared to the results of the present
analysis.
\section{
Models for the proton structure functions}
\subsection{Common set of data included in the fitting procedure}
The choice of the data set used in the fits may have crucial
consequences when calculating the derivatives of the SF,
especially when this quantity is to be discussed near the
experimental kinematical limit or/and near the theoretical
validity limit. For these reasons, a unique set including the most
recent data has been used in new fits of three models of the proton SF
we choose to discuss.

These updated data are listed and referenced in Table 1. The
complete available experimental kinematical range is taken into
account for $Q^2$ (\ie $0.\leq Q^2$ $ \leq 30000$ \g2) and $x$ (\ie $6.\
10^{-7}\leq x\leq 0.75$), with the lower limit of the center of mass energy
for ${\gamma^*p}$ scattering $W\geq 3$ GeV.

\medskip
{\small
\begin{displaymath}
\begin{array}{|cc|c|c|c|}
\hline
{\rm Observable}  & & & \chi^2 & \chi^2 \\
{\rm Exp.-year\ of\ pub., }&{\rm \ Ref} & {\rm Nb\,points}&
{\rm model~\cite{djp}} & {\rm model~\cite{dlm} } \\
\hline \hline
 F_2^p                      & & &  &  \\
{\rm H1-1995}      \ \ & \cite{ah95}&  93 &  76.1 & 70.7 \\
{\rm H1-1996}      \ \ & \cite{ai96}& 129 &  90.9 & 68.9 \\
{\rm H1-1997}      \ \ & \cite{ad97}&  44 &  53.0 & 53.5 \\
{\rm H1-1999}      \ \ & \cite{ad99}& 130 & 237.9 & 111.0 \\
{\rm ZEUS-1996}    \ \ & \cite{de96}& 188 & 259.9 & 261.5 \\
{\rm ZEUS-1997}    \ \ & \cite{br97}&  34 &  12.2 & 17.3 \\
{\rm ZEUS-1999}    \ \ & \cite{br99}&  44 &  26.7 & 38.5 \\
{\rm ZEUS-2000}    \ \ & \cite{br00}&  70 & 139.6 & 79.3  \\
{\rm NMC-1997}     \ \ & \cite{ar97}& 156 & 285.5 & 161.1 \\
{\rm E665-1996}    \ \ & \cite{ad96}&  91 & 109.7 & 99.5 \\
{\rm SLAC-1990/92} \ \ & \cite{slac}& 136 & 193.6 & 107.3 \\
{\rm BCDMS-1989}   \ \ & \cite{be89}& 175 & 274.8 & 269.9 \\
\hline
\sigma_{tot}^{\gamma ,p} \ \ &  & & & \\
{\rm 1975/78;ZEUS-1994;H1-1995} & \cite{ca75} & 99 &  148.6 & 122.7 \\
\hline
{\rm Total}              &   &1389 &1907.6 &1461.2  \\
\hline \hline
\chi^2\ / \ d.o.f.       &   &   &1.38 &1.07  \\
\hline
\end{array}
\end{displaymath}
}

\smallskip
\noindent {\bf Table 1.} Observables sets used in the fitting
procedure. Also shown are the distributions of the partial \x2 for
each subset of data used in our new fits with the parameters listed
below in Tables 2-3, for the two updated models~\cite{djp,dlm}.

\medskip
Generally speaking, we must note a good agreement between both
HERA experiments, H1 and ZEUS, and the fixed target ones.
Following the suggestion from~\cite{ad99}, some data from
~\cite{ai96} are considered as obsolete and superseded. These
correspond to ($Q^2\ge 250$ \g2 , for all $x$), ($Q^2= 200$ \g2 ,
for $x<0.1$) and ($Q^2= 150$ \g2 , for $x<0.01$).

\medskip
\subsection{ Parametrizations of the structure function}
\subsubsection{The ALLM model}
The so-called ALLM parametrization~\cite{AL-97} for describing the
total $\gamma^*p$ cross section, connected to the proton SF, has
been updated in 1997 to account for the set of available data in
the whole range studied so far: $0<Q^2<$ 5000 \g2, and $3.
10^{-6}<x<0.85$. Here, to make the comparison with the other
results more meaningful, we extend the $x, Q^2$ ranges over all
the data of Table 1 and perform a new fit of the 23 parameters of
this model. We find again quite a good agreement which can be
characterized by \xi2=1.06

\smallskip
\subsubsection{The Lyon-Kiev-Padova model revisited}
We emphasize that this LKP model was created for interpolating
between soft and hard dynamics, combining Regge behavior and the
high $Q^2$ asymptotics of the DGLAP evolution equation, for the
low$-x$ range, with a small number of free parameters (8,
in~\cite{djp}).

To improve the quality of the fit on all data, we accept to lose
the simplicity of the original model by modifying here the
"large" $x$ extension which is no more borrowed from the
relatively simple and rather efficient model~\cite{CKMT} but,
instead, we follow the more sophisticated approach from~\cite{dlm}.
Explicitly, the SF has been rewritten \be F_{2}(x,Q^2) =\
F_{2}^{(S,0)}(x,Q^2) \cdot (1-x)^{P(Q^2)} +\
F_{2}^{(NS,0)}(x,Q^2)\cdot (1-x)^{R(Q^2)}\ , \ee where the Pomeron
contribution (singlet $F_{2}^{(S,0)}$) and the effective secondary
Reggeon component (non-singlet $F_{2}^{(NS,0)}$) have here the
same expressions as in~\cite{djp} at low$-x$, and with the
following $Q^2$-dependent exponents of the large$-x$ factors \be
 P(Q^2)=\ p_\infty+{p_0-p_\infty\over 1+Q^2/Q^2_p}\ ,
\quad R(Q^2)=\ r_\infty+{r_0-r_\infty\over 1+Q^2/Q^2_r}\ , \ee
where $p_\infty, p_0, r_\infty, r_0, Q_p, Q_r$ are thus additional parameters
in the present version of the LKP model . In
definitive, taking into account the large $x$ in practice doubles
the parameters number, but it is the price of such an extension.
The details on the performances of this extended model are given
in Table 1 in terms of the \x2 for each subset of data, yielding
\xi2=1.38. The parameters we find are listed in the Table 2, with their meaning
explained in~\cite{djp} and above.
\medskip
{\small
\begin{center}
\begin{tabular}{|c||c|c|}
\hline
Parameters   & previous work~\cite{djp}& present work   \\
\hline \hline
\strut low $x-$terms  & & \\
\hline
$A$                      &    $0.1470      $ & 0.1190 \\
$a$ (GeV$^2$)            &    $0.2607      $ & 0.2300 \\
$\epsilon$               & 0.08 (fixed)      & 0.0895        \\
$\gamma_2$               &    $0.0200      $ & 0.0221 \\
$Q_0^2$ (GeV$^2$)        &    $0.1675      $ & 0.1946 \\
$Q_1^2$ (GeV$^2$)        &    $1174.       $ & 7800.  \\
$B$                      &    $0.7575      $ & 1.6409 \\
$b$ (GeV$^2$)            &    $0.4278      $ & 1.46 \\
$\alpha_r$               &    $0.5241      $ & 0.48 (fixed) \\
\hline
\strut large $x-$terms   &(fixed)          & \\
\hline
$Q^2_p$ (GeV$^2$)        & $-$             & 1.1180  \\
$p_0$                    & $-$             & $\approx 0.$ \\
$p_\infty$               & $-$             & 15.093  \\
$Q^2_r$ (GeV$^2$)        & $-$             & 12.563  \\
$r_0$                    & $-$             & 2.394  \\
$r_\infty$               & $-$             & 3.728  \\
\hline
\end{tabular}
\end {center}
}
\smallskip
\noindent {\bf Table 2.} Parameters used in the previous and present versions of
the LKP model. The
non-fitted values are: $\gamma_1=2.4$ (suggested by QCD) for both
versions, $\epsilon=0.08$ (from~\cite{eps}) and the large $x$
exponents (from~\cite{CKMT}) for the previous work only.

\medskip

When the intercept $\alpha_r$ of the $f$-Reggeon is allowed to be free
we obtained a better $\chi^{2} (\approx 1.22)$. However the
resulting value of $\alpha_{r}\approx 0.24$ is totally
unacceptable from a physical point of view. For that reason, we fixed it,
choosing $\alpha_{r}=0.48$, a "mean" value
used in the models with a supercritical Pomeron.
\smallskip

\subsubsection{Soft dipole Pomeron model}

The refitted parameters of this model are given in Table~3 together
with the old ones published in \cite{dlm} (the same notations are used).
The most interesting property of this model developed in \cite{dlm} is the
intercept of Pomeron. It does not depend on $Q^{2}$ and moreover it is
equal to one. Nevertheless due to an interference of two Pomeron
components, the model describes well all observed properties of the proton
structure function, including a growth with $Q^{2}$ of the effective
Pomeron intercept (see for details \cite{dlm} and \cite{dlm1}). The first
Pomeron component, dominating at small $x$, leads to
$F_{2}(x,Q^{2})\propto \ell n(1/x)$ at $x\to 0$ while the second one
gives asymptotically a constant and negative contribution to $F_{2}$
playing an important role in the whole kinematic region where the data on
the proton SF data are available.

\medskip
\begin{center}
 {\small
\begin{tabular}{|c||c|c|}
\hline
 Parameters  & previous work~\cite{dlm}& present refit   \\
\hline \hline
\strut $P_1$-term  & & \\
\hline
$\mu$                     & .10000E+01 (fixed)&.10000E+01 (fixed) \\
$\alpha_{\cal P}(0)$      & .10000E+01 (fixed)&.10000E+01 (fixed) \\
$g_1$ (mb$)$              & .21898E-01        &  .22198E-01 \\
$Q_1^2$ (GeV$^2)$         & .15400E+02        &  .70711E+01 \\
$Q_{1d}^2$ (GeV$^2)$      & .17852E+01        &  .14774E+01 \\
$Q_{1b}^2$ (GeV$^2)$      & .33435E+01        &  .67975E+01 \\
$d_{1\infty}$             & .13301E+01        &  .12601E+01 \\
$d_{10}$                  & .14370E+02        &  .66975E+01 \\
$b_{1\infty}$             & .21804E+01        &  .28712E+01 \\
$b_{10}$                  & .42596E+01        & -.20279E+01 (fixed)\\
\hline
 $P_2$-term               &         & \\
\hline
$g_2$ (mb)                & -.99050E-01       & -.10176E+00  \\
$Q_2^2$ (GeV$^2)$         & .34002E+02        &  .13748E+02  \\
$Q_{2d}^2$ (GeV$^2)$      &.12327E+01         &  .15954E+01  \\
$Q_{2b}^2$ (GeV$^2)$      & .20702E-01        &  .80605E+01  \\
$d_{2\infty}-d_{1\infty}$ & .00000E+00 (fixed)&  .00000E+00 (fixed) \\
$d_{20}$                  & .22607E+02        &  .64794E+01  \\
$b_{2\infty}$             &.24686E+01         &  .34510E+01  \\
$b_{20}$                  & .17023E+03        &  .12922E+01  \\
\hline
 $F$-term  & & \\
 \hline
$\alpha_f(0)$             & .80400E+00 (fixed)&  .80400E+00 (fixed) \\
$g_f$ (mb)                &.29065E+00         &  .29405E+00  \\
$Q_f^2$ (GeV$^2)$         & .29044E+02        &  .10182E+02  \\
$Q_{fd}^2$ (GeV$^2)$      & .54462E+00        &  .70413E+00  \\
$Q_{fb}^2$ (GeV$^2)$      & .20656E+01        &  .84803E+00  \\
$d_{f\infty}$             &.13554E+01         &  .13149E+01  \\
$d_{f0}$                  & .75127E+02        &  .19746E+02  \\
$b_{f\infty}$             & .27239E+01        &  .33642E+01  \\
$b_{f0}$                  & .64713E+00        & -.27968E+01 (fixed)  \\
\hline
\end{tabular}
}
\end{center}
\smallskip
\noindent {\bf Table 3.} Parameters refitted in the soft Dipole Pomeron
model of~\cite{dlm}, the original values are also quoted.

Analyzing the properties of the model we found that it is not necessary to
restrict the parameters $b_{10}, b_{20}$ and $b_{f}$ (which influence the
behaviour of $F_{2}$ at $x\to 1$) to positive values. Indeed, there
are two domains when $x\approx 1$. In the first one, $Q^{2}$ is large,
$Q^{2}\gg W^{2}-m^{2}$, the $x$-behaviour of $F_{2}$
is controlled by $b_{i\infty}$ rather than by $b_{i0}$. In the second
domain $W^{2}$ is small, $W^{2}-m^{2}\approx 0$, it is a region
where the Regge approach cannot be applied.

Thus, we allowed the negative values for
$b_{10}, b_{20}$ and $b_{f0}$ and found out a new set of parameters
which, though noticeably different from the old one, leads to a very good
$\chi^{2} \approx 1.07$.

\smallskip

\bigskip
\section{ Extracting the local slopes $B_{x}(x,Q^{2})$ from the
SF data.}
As noted in the introduction, the $x-$slope is a precious tool to
settle if, either yes or no, a damping of the HERA effect does
exist. Actually, we need a set of experimental data on
$B_{x}(x,Q^{2})$, at fixed and not too high $x$ versus
sufficiently high $Q^{2}$. Unfortunately, because the $x-$slope is
not a measurable observable it should be extracted, when possible,
from the available data on the SF.

For each given $Q^{2}$, we have an insufficient number of $x$
values for which SF data are available, to perform an analysis
based on independent $x-$ bins and to extract $x$-slopes with a
good accuracy. We adapt the so-called method of "overlapping
bins"~\cite{konlen}, previously intended for analyzing the local
nuclear slope of the first diffraction cone in $pp$ and $\bar pp$
elastic scattering.

\smallskip

Provided that the SF has been measured for a given $Q^{2}$ at $N$
$x$-points lying in some interval $[x_{min},x_{max}]$, we adopt
the following procedure. First, we divide this interval into
subintervals or elementary "bins" (with $n_{b}$ measurements in
each of them, assumed for simplicity to be the same for all bins).
When the first bin is chosen, the second bin is obtained from the
first by shifting only one point of measurement (of course one
could shift bins by any number of points less or equal $n_{b}$,
the shift of one point is the minimal one giving rise to the
maximal number of overlapping bins). The third bin is obtained
from the second one by the shift of one point \etc Thus, we define
$N-n_{b}+1$ overlapping bins for a given $Q^2$. For each ($k$-th)
bin, $n_{b}$ must be large enough and its width (in $x$) small
enough to allow fitting the SF with the simplest form directly
involving the $x-$slope
\begin{equation}\label{6}
  F_{2}(x)=A\left (\frac{1}{x}\right)^{B}\ ,\quad ({\rm for\ a\ given }\
  Q^2)\ .
\end{equation}
The parameter $B$ represents the value of the $x-$slope
$B\left(<x>_{k},Q^{2}\right)$, "measured", at $Q^2$ and at the
"weighted average" $<x>_{k}$ defined in the $k$-th bin as (see
\cite{br99})
\begin{equation}\label{7}
  <x>_k=exp\bigg (-\frac{\sum\frac{\ln x_{i}}{\Delta y_{i}}}
  {\sum\frac{1}{\Delta y_{i}}}\bigg)\ ,\quad
  k\in{[1,N-n_{b}+1]}\      ,
\end{equation}
where $x_{i}$ is the value of $x$ at which the structure function
$y_{i}$ is measured with the uncertainty $\Delta y_{i}$; the
summations run over all data points, $i=1,2,...,n_{b}$ of the bin.
This yield the "experimental" values of $B_{x}(x,Q^{2})$ with the
corresponding standard errors determined in the fit of (6) to the
data. Then the procedure is to be repeated for all bins and
ultimately for the other $Q^{2}$'s at which the SF have been
measured.

\smallskip

The next step in extracting and analyzing $B_{x}(x,Q^{2})$ is
the determination of the slopes at fixed $x$ as function of $Q^{2}$,
making use of results of the first step. As a rule, the sets of
$<x>_{k}$, at different $Q^{2}$, do not coincide. So in order to get the
$x-$slope at fixed $x$ and at different $Q^{2}$ we
interpolate (or extrapolate but not far from the $x$-interval under
consideration) already extracted $B_{x}$ at the given $Q^{2}$ to
the chosen $x$. It can be made assuming a linear $x-$dependence
of $B_{x}$.

The above mentioned method of overlapping bins is applied to the whole
available data set. The HERA data (of ZEUS and H1) are analyzed all
together~[18-25] but independently (separately) from older data of the
other experimental Collaborations~[26-30] because there are large gaps
between the $x$-values for these two groups of data (remind that our aim
is to extract the local $B_{x}$ rather than averaged ones).

The resulting values of $B_{x}(x,Q^{2})$ are shown in the Figs~.2 (a-e),
where they are compared to the predictions of theoretical models. The
results of the interpolation for $x=0.005, 0.01, 0.05 $ and 0.08 are
presented in the Fig.~3.

We would like to comment some "technical" points in our analysis
and results.
\begin{itemize}
\item[i)]
In order to keep its local character to the $x$-slope and to
obtain a maximal possible number of "measurements", we have
considered only the cases with four and five points in each
elementary bin. One can see from the presented figures a quite
weak dependence of the results on a change in the width of
elementary bin.
\item[ii)]
In spite of a high accuracy of the recent data from HERA, the
dispersion of the SF-values strongly influences the resulting
values of $B_{x}$ and of its errors. For some bins, for example,
we were unable to fit (6) to the data with $\chi^{2}\leq 1$, so we
could not obtain in all cases reasonable errors. Nevertheless we
show these extracted values (only with $\chi^{2}\leq 3$)in the
figures and use them for interpolation to the $x$ under interest
because even with these points the extracted set of data for
$B_{x}(x,Q^{2})$ at fixed $x$ is quite poor. Of course, this
reduces the reliability of our results, but only slightly because
the number of "bad" bins remains small.
\item[iii)]
"Experimental" values of $B_{x}$ at fixed $x$ shown in Fig.~3 are
obtained by a linear interpolation within the two subsets of local
$x$-slopes, extracted from the HERA and from the fixed target
measurements of SF.
\item[iv)]
One can see in Fig.~3 that several points deviate strongly from
the groups constituted by the other points. This is due to the
strong influence of the points (of $F_{2}$ as well as of $B_{x}$)
which are at the ends of the $x$-bins and which also "fall out of
a common line" (see also item ii) above). To solve this problem, a
possible way would be to exclude some of them from the analysis; another
way would be to enlarge the number of points in an elementary bin
losing, however, the local character of the extracted $x-$slope. A
more detailed analysis of the available data related to more
numerous measurements of the primary observable, $F_{2}(x,Q^{2})$
would  be necessary to obtain a better set of data for $B_{x}(x,Q^{2})$.
\end{itemize}

\newpage
\noindent
\begin{minipage}[th]{7.52cm}
\begin{center}
\includegraphics*[scale=0.44]{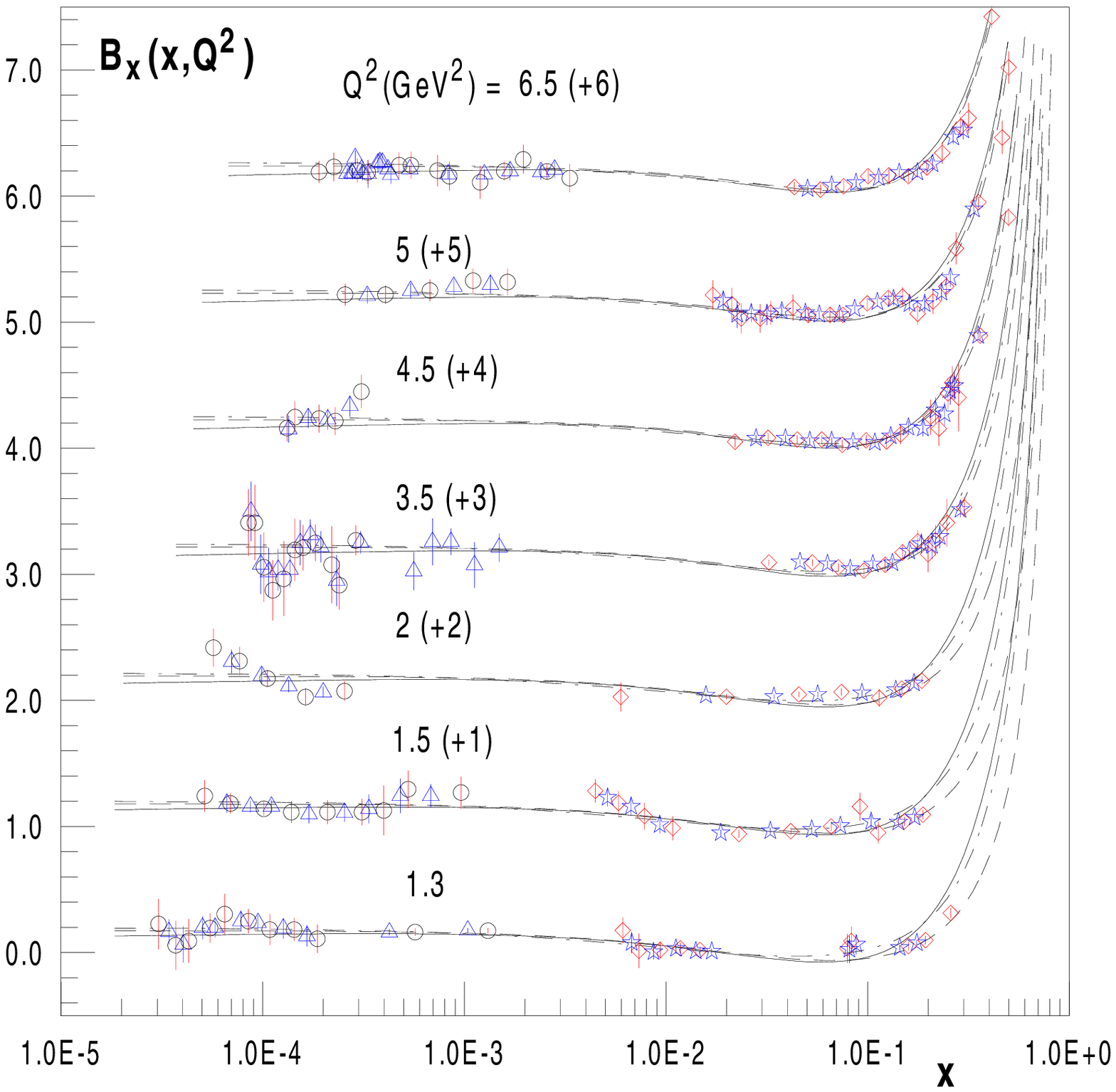}
 (a)
\end{center}
\end{minipage}
\hskip 1.1cm
\begin{minipage}[th]{7.52cm}
Fig.~2 (a-e). The $x-$slopes $B_{x}(x,Q^{2})$ at fixed
$Q^{2}$ and versus $x$, extracted from the experimental data on the
structure function by the method of overlapping bins. Triangles
($n_{b}=5$) and circles ($n_{b}=4$) correspond to the slopes extracted
from the HERA data. Stars ($n_{b}=5$) and diamonds ($n_{b}=4$) are the
slopes extracted from fixed target data. Curves are predictions in the DP
(solid line), ALLM (dashed line) and LKP (dashed-dotted line) models.
\end{minipage}

\noindent
\begin{minipage}[th]{7.52cm}
\begin{center}
\includegraphics*[scale=0.44]{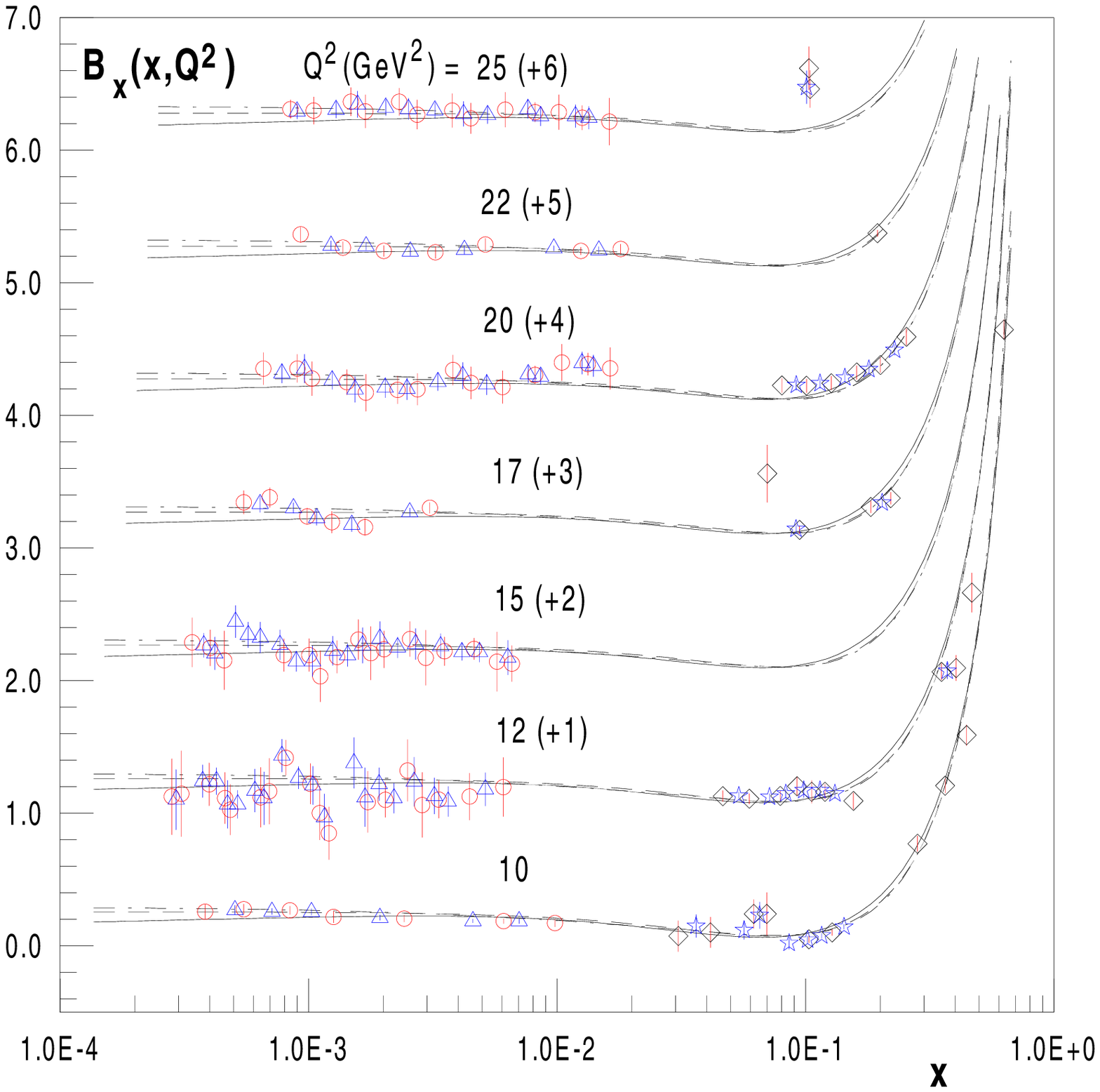}
 (b)
\end{center}
\end{minipage}
\hskip 1.1cm
\begin{minipage}[th]{7.52cm}
\begin{center}
\includegraphics*[scale=0.44]{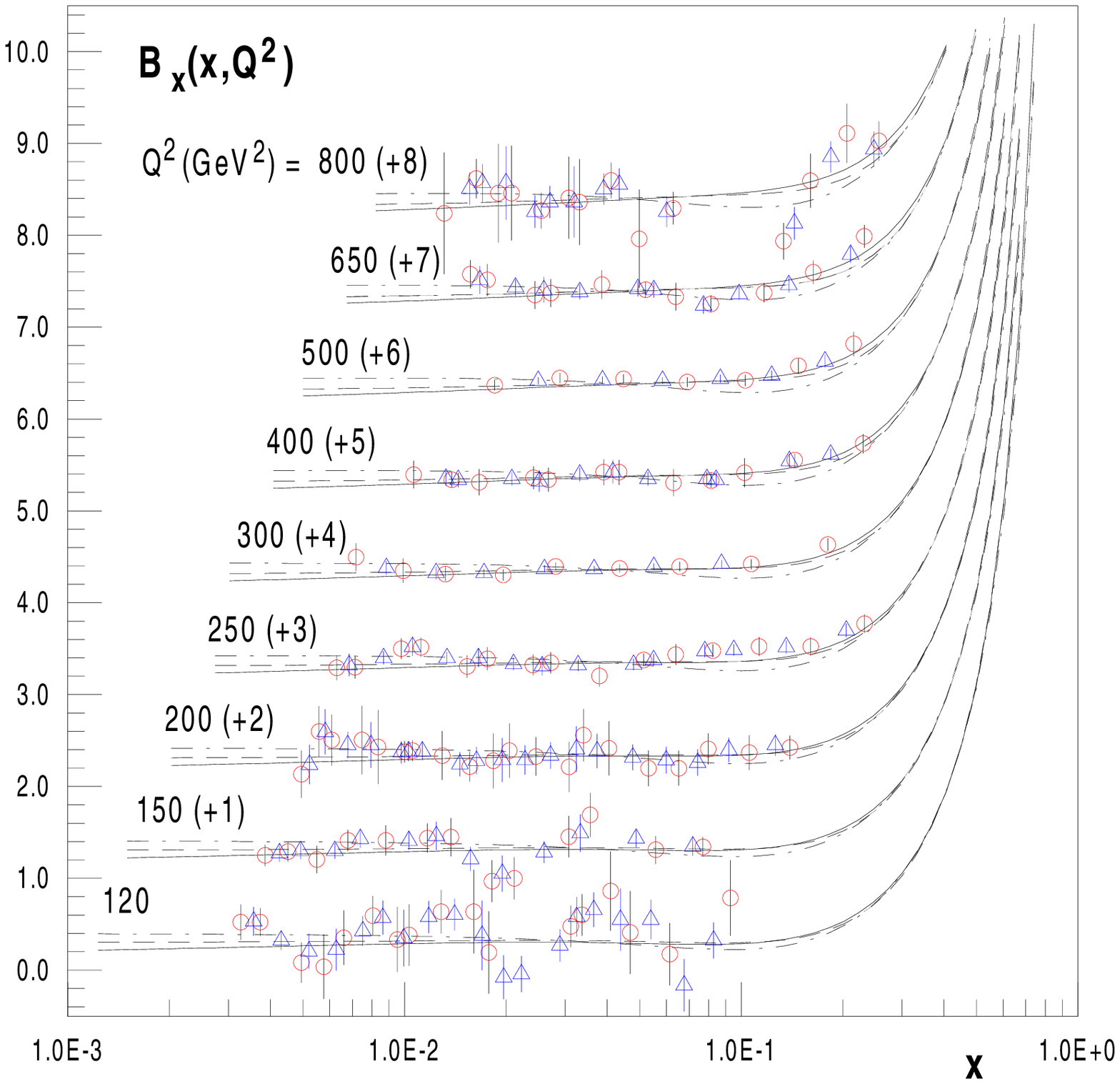}
 (d)
\end{center}
\end{minipage}

\noindent
\begin{minipage}[th]{7.52cm}
\begin{center}
\includegraphics*[scale=0.44]{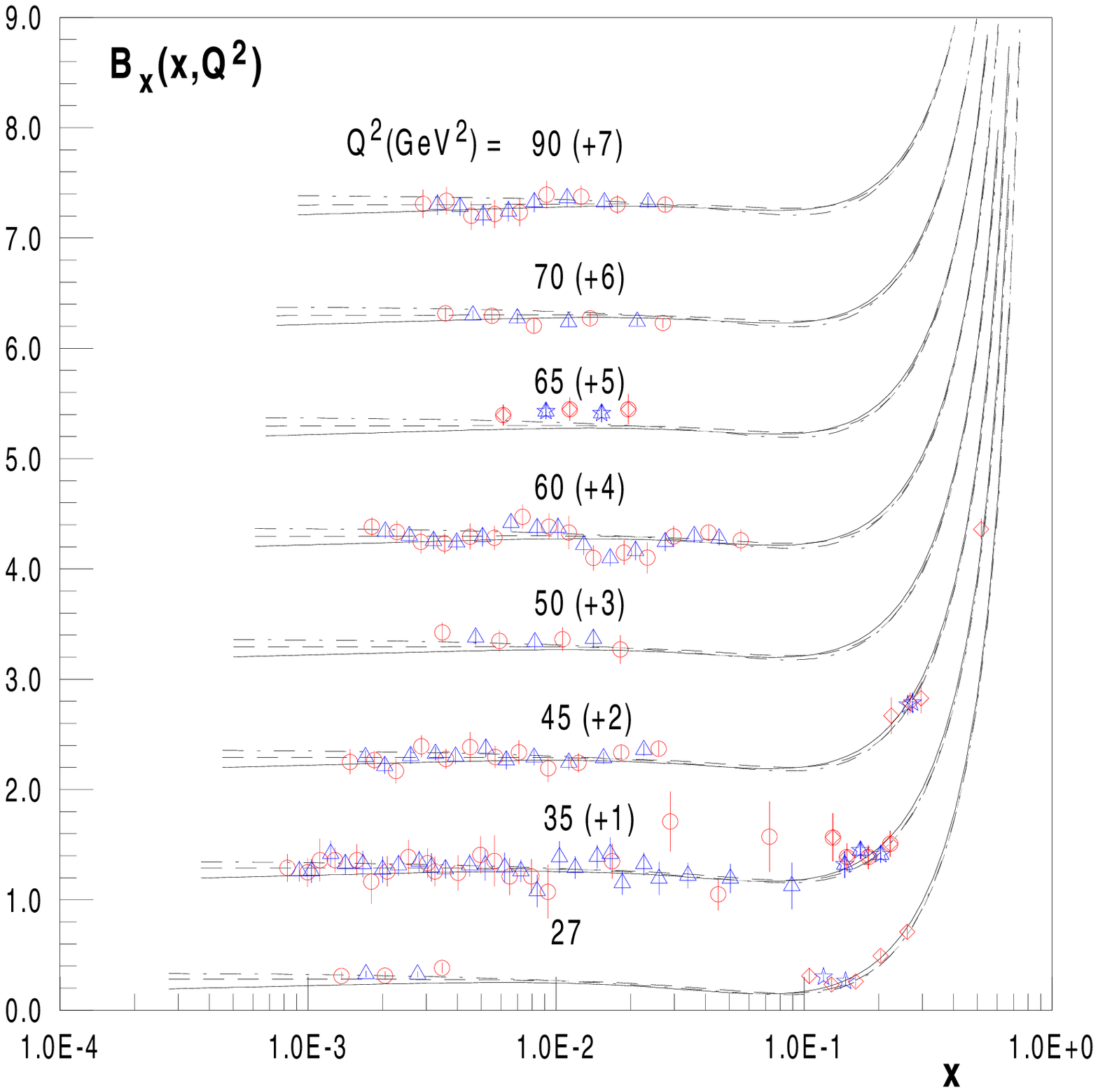}
 (c)
\end{center}
\end{minipage}
\hskip 1.1cm
\begin{minipage}[th]{7.52cm}
\begin{center}
\includegraphics*[scale=0.44]{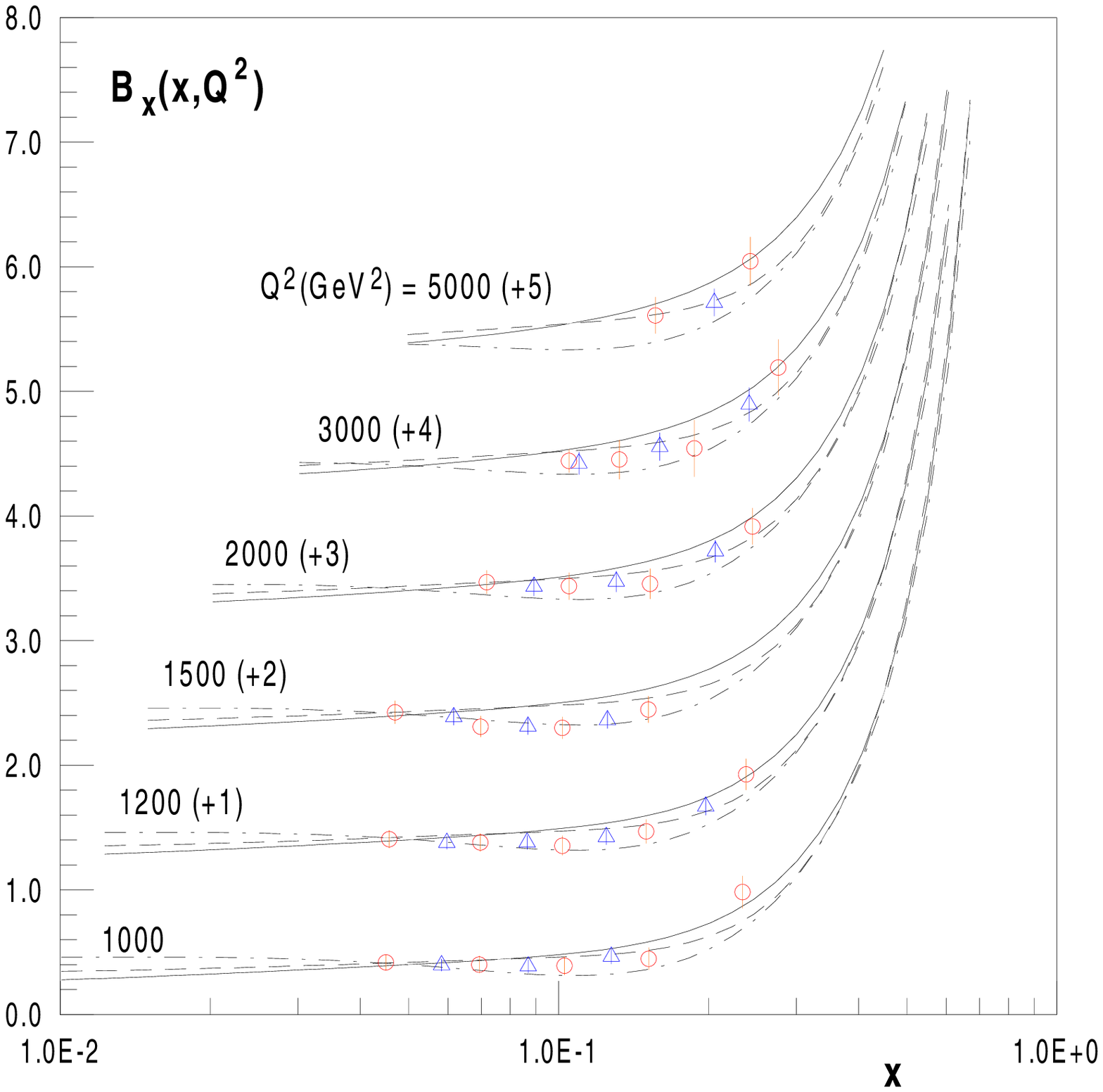}
 (e)
\end{center}
\end{minipage}

\bigskip

\begin{center}
\includegraphics*[scale=0.85]{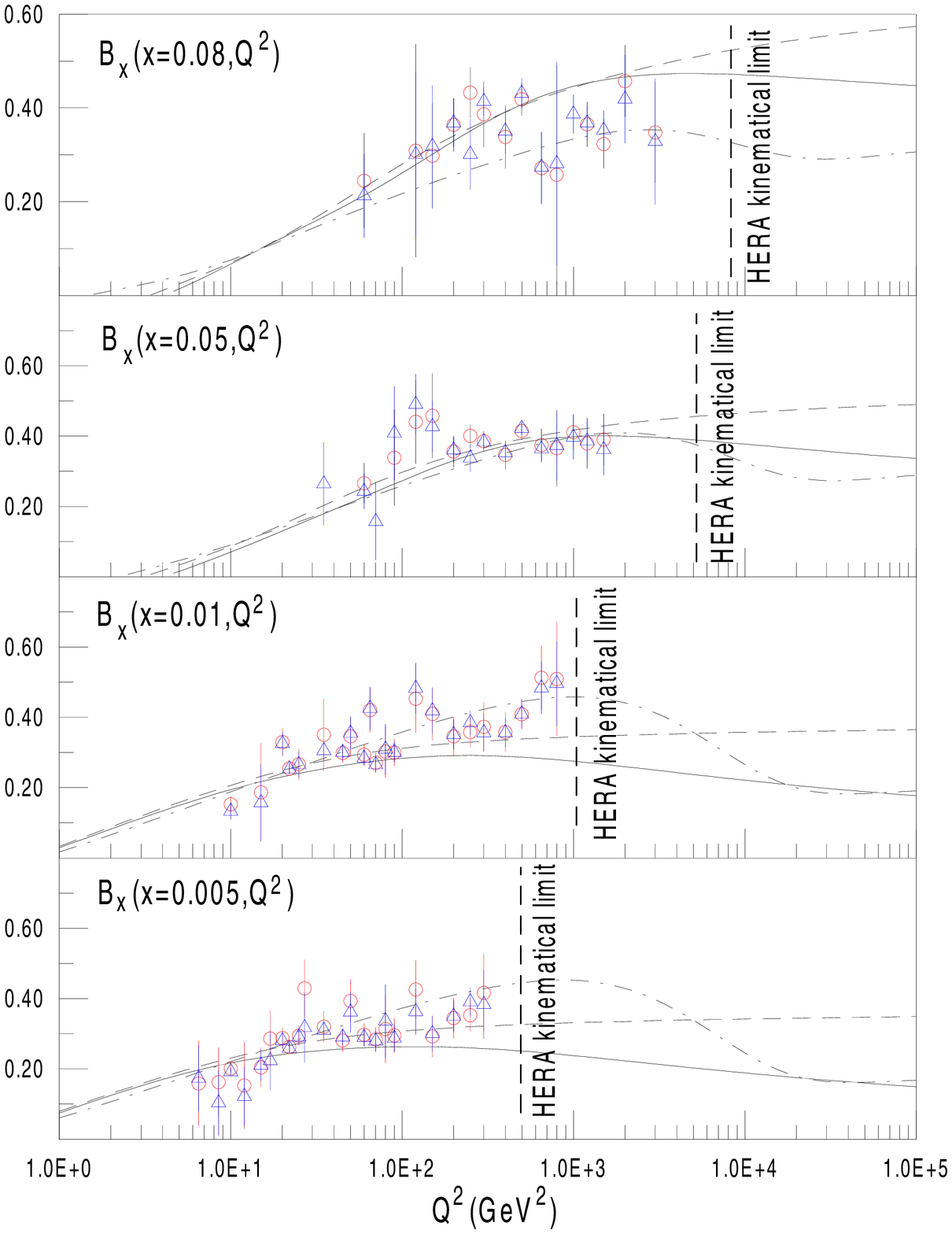}
\end{center}
Fig.~3. The slopes $B_x$ interpolated to some $x$-s as function of
$Q^2$. The vertical dotted lines show the HERA kinematical limits
for given $x$. See Fig.~2.

\section{Results and discussion}


The calculated $x-$slopes within the three models are in good agreement
with the values extracted from the experimental SF via the present
analysis (in the data bars, see Figs~.2), except at largest $x$ where the
predictions are slightly higher than experimental points.

Except for the highest $Q^2$, in Fig.~2(e), and for the lowest $x$
in all figures, the theoretical curves almost coincide. To that
respect, the slope does not appear as a good criterion for
discriminating among the various models. However, we found that
the three models predict noticeably different $x-$slopes when $x$
is very small, outside reported experimental range (the difference
growing when $x$ decreases)~\footnote{ In fact, when $x\to 0$,
only the DP model is in agreement with the implication of the
Froissart-Martin bound for the $\gamma^{*} p$ cross-section
concerning the nullity of $B_x(x\to 0,Q^2)$.}. For example, we
find $B_x(x=10^{-4},Q^2=100 \mbox{GeV}^{2})\approx 0.2$ for DP,
$\approx 0.4 $ for LKP, and $\approx 0.3 $ for ALLM.

Thus, an extension of the $x,Q^2$ range for $B_x$ towards small
$x$, requiring much effort in measuring and analyzing the FS,
would be very interesting to seek the experimental evidence of the
validity (or non validity) of the Froissart-Martin bound in deep
inelastic scattering by ensuring whether (or not) $B_{x}(x\to
0,Q^{2})\to 0$. Furthermore, exploration of this kinematical
region allows to select a model upon experimental grounds.

\smallskip

The damping effect is characterized by a slowing down of the
$x-$slope growth, when $Q^2$ increases for any $x$, resulting in a
turn-over in the behavior of th $x-$slope.

From the theoretical point of view, we recall that we tested only three
models~\cite{AL-97,djp,dlm}, which assume quite different asymptotic
behaviors of the SF and consequently of $B_x(x\ll 1, Q^2\to\infty)$ (for
more details see~\cite{dlm}). The Fig.~3 shows the $x-$slope calculated in
these models and plotted versus $Q^2$, for some $x-$ values. Only LKP and
DP models do predict~\footnote{ No correlation is found between the
occurrence of a bump and the asymptotic behavior, depending on the
parametrization.} a bump in the curve for $Q^2$ above 100 \g2, ALLM does
not~: consequently, we cannot rely on theoretical predictions alone due to
that model dependence.

From the experimental point of view, the figure shows data points
resulting from our analysis, seeming to correspond to a maximum.
However, the most interesting part of the $Q^2$ range to explore
in the future (because situated in the downhill of the predicted
$x-$slope and allowing to conclude unambiguously) lies above the
actual measurements and not too far from the actual kinematical
limit (at least for HERA facilities).

We conclude that a deeper theoretical knowledge of the high
energy positron-proton interaction or/and more precise
measurements and analysis, opened in higher $Q^2$ range, are
needed to confirm or to disprove without ambiguity a bump
structure in the behavior of $B_{x}(x,Q^{2})$ at small fixed $x$.

\bigskip

{\it Acknowledgements} We have the pleasure to thank L. Jenkovszky
for interesting discussions and M. Giffon for a critical reading of the
manuscript. E.M. is grateful to Univercit\'e
Claude Bernard and IPNL for the kind hospitality and financial
support extended to him during this work.

\bigskip

\end{document}